\documentclass[aps,prb,twocolumn,groupedaddress]{revtex4}

\usepackage{graphicx}

\begin{document}

\title{Entanglement in a Noninteracting Mesoscopic Structure}

\author{A.V.\ Lebedev$^{\, a}$, G.B.\ Lesovik$^{\, a}$,
and G.\ Blatter$^{\, b}$}
\affiliation{$^{a}$L.D.\ Landau Institute for Theoretical Physics RAS,
117940 Moscow, Russia}

\affiliation{$^{b}$Theoretische Physik, ETH-H\"onggerberg, CH-8093
Z\"urich, Switzerland}

\date{\today}

\begin{abstract}
   We study the time dependent electron-electron and
   electron-hole correlations in a mesoscopic device
   which is splitting an incident current of free fermions
   into two spatially separated particle streams. We
   analyze the appearance of entanglement as manifested
   in a Bell inequality test and discuss
   its origin in terms of local spin-singlet correlations
   already present in the initial channel and the action
   of post-selection during the Bell type measurement.
   The time window over which the Bell inequality is
   violated is determined in the tunneling limit and for the
   general situation with arbitrary transparencies.
   We compare our results with alternative Bell inequality
   tests based on coincidence probabilities.
\end{abstract}

\maketitle

\section{Introduction}

Quantum entanglement of electronic degrees of freedom in
mesoscopic devices has attracted a lot of interest recently.
The first proposal to probe localized entangled electrons
through transport and noise measurements \cite{loss}
soon lead to specific structures which generate spatially
separated streams of entangled particles \cite{ent_sc,ent_qd}.
One class of devices makes use of a superconducting source
emitting Cooper pairs into a normal-metal structure with
two leads in a fork geometry: entanglement has its origin
in the attractive interaction binding the electrons into
Cooper pairs, while the spatial separation of correlated
electrons is arranged for by suitable `filters'
\cite{ent_sc}. Another class of devices makes use of
Coulomb interactions in confined geometries \cite{ent_qd}.
All of these proposals involve electronic spins as
the entangled quantum degrees of freedom; an alternative
scheme has been pointed out by Samuelsson {\it et al.}
\cite{samuelsson_03} who propose a setup where real space
orbital degrees of freedom become entangled. Besides these
proposals for the generation of spatially separated
entangled pairs, the implementation of Bell inequality
tests probing their entanglement has been discussed in
detail \cite{kawabata_01,chtchelkatchev_02,samuelsson_03}.
The combination of sources for the creation and methods
for testing the correlations of entangled particle streams
are first steps towards establishing this quantum resource
for solid state based quantum information technology.
\begin{figure} [h]
   \includegraphics[scale=0.45]{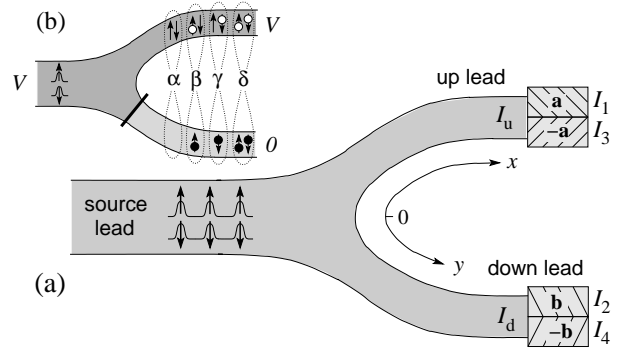}
   \caption[]{(a) Mesoscopic normal-metal structure in a fork geometry
   generating two streams of spin-correlated electrons in the two arms
   of the fork. The Bell type setup detects spin-currents $I_i$,
   $i=1,3$, projected onto the directions $\pm{\bf a}$ in the upper arm
   and correlates them with spin-currents $I_j$, $j=2,4$, projected onto
   the directions $\pm{\bf b}$ in the lower arm. (b) In the tunneling
   limit (with a small transparency $T_\mathrm{d} \ll 1$ in the `down'
   arm) a fraction of spin-entangled electrons is split into the two
   arms (components $\beta$ and $\gamma$); their correlations
   can be efficiently measured in a Bell type setup sensitive to hole
   (electron) currents in the `up' (down) lead.}
   \label{fig:fork}
\end{figure}

An interesting proposal has recently been made by Beenakker
{\it et al.} \cite{beenakker_03} (see also Ref.\
\onlinecite{beenakker_03b} and the note \onlinecite{note}):
using a two-channel quantum Hall device with a beam
splitter, they suggest a setup generating two streams of
entangled electron-hole pairs and confirm the presence of
correlations through a Bell inequality (BI) test. A crucial
difference to previous proposals is the absence of interactions
generating the entanglement (see also Refs.\
\onlinecite{note,fazio_03,ssb_04}).
In this paper, we investigate a similar setup involving
a mesoscopic normal-metal structure arranged in a fork
geometry and generating two streams of correlated electrons
in the two arms of the fork, see Fig.\ \ref{fig:fork}. In
our formulation of the Bell inequalities we take special
care to use only directly measurable observables. We
find that a Bell inequality test involving correlations
between the electronic spin-currents in spatially
separated leads exhibits violation at short time scales $\tau <
\tau_{\rm\scriptscriptstyle BI}$, with $\tau_{\rm
\scriptscriptstyle BI}$ given by the single particle correlation
time $\tau_V = \hbar /eV$. We show, that the time $\tau_{\rm
\scriptscriptstyle BI}$ can be considerably extended in the
tunneling limit, where the propagation into one of the arms (we
choose the `down' lead in Fig.\ \ref{fig:fork}) is strongly
reduced by a small transparency $T_\mathrm{d}$. This small
transparency can be exploited by a suitable choice of the
observables in the Bell inequality test: going over to hole
currents in the well conducting `up' lead, we find the long
violation time $\tau_{\rm \scriptscriptstyle BI}= \tau_V
/\sqrt{T_\mathrm{d}}$. We also analyze an alternative Bell
inequality based on coincidence probabilities derived from
electron-hole {\it number} correlators and find it violated
on even longer times $\tau_{\rm \scriptscriptstyle BI}= \tau_V
/T_\mathrm{d}$. When tracing the origin of these BI
violations, we find that the fermionic statistics already
enforces the injection of spin-singlet correlated electron
pairs with correlations extending over the distance $v_{\rm
\scriptscriptstyle F} \tau_V$ with $v_{\rm \scriptscriptstyle F}$
the Fermi velocity. In addition, the splitting of this pair
and a subsequent post-selection \cite{shih_88,ssb_04} in the
Bell type measurement are crucial steps in rendering the original
spin-entanglement amenable to observation and potentially
useful as a resource of entangled quantum degrees of freedom.

In the following, we first analyze (Sec.\ II) the two-particle
density matrix of the injected particle stream and find it
to be locally singlet correlated; the extension to the
scattered state behind the beam splitter demonstrates
that this entanglement is preserved. We proceed with a
discussion of Bell inequality measurements (Sec.\ III)
and compare two types of Bell parameters, one based on
current cross-correlators and a second one starting from
coincidence probabilities, i.e., particle number
correlators. We then turn to the actual calculation
of the finite time spin-current correlators
and their combination into the Bell parameter (Sec.\ IV).
Results are presented for Bell inequalities involving
electron-electron (general case) and electron-hole
(tunneling limit) current correlations as well as
for the corresponding expressions based on coincidence
probabilities; we demonstrate that the violation of the
Bell inequality depends sensitively on the choice of
observables. Finally, we present our conclusions
in Sec.\ V.

\section{Two-particle correlations}

In order to analyze the properties of the injected electrons,
we determine the two-particle density matrix (or pair correlation
function) within the source lead,
\begin{eqnarray}
   &&
   g_{\vec\sigma}(x,y)=
   {\rm Tr}\bigl(\hat\rho\, \hat\Psi_{\sigma_{1}}^\dagger(x)
   \hat\Psi_{\sigma_{2}}^\dagger(y)
   \hat\Psi_{\sigma_{3}}(y)
   \hat\Psi_{\sigma_{4}}(x)\bigr)
   \label{pair_s}\\
   &&
   =\langle\sigma_1|\sigma_4\rangle\langle\sigma_2|\sigma_3\rangle
   \, G^2(0)-
   \langle\sigma_1|\sigma_3\rangle\langle\sigma_2|\sigma_4\rangle
   |G(x-y)|^2,
   \nonumber
\end{eqnarray}
where we have introduced the one-particle orbital correlator
$G(x)\equiv\langle\hat\Psi^\dagger(x) \hat\Psi(0) \rangle$, with
$\hat\Psi(x)$ the field operator describing spinless electrons in
the source lead and $\sigma_i$ are spin indices. We ignore
contributions from backscattering originating from the splitter
and describe the source lead as a ballistic wire connecting two
reservoirs with Fermi levels shifted by the voltage bias $eV$. The
correlator can be separated into an equilibrium- and an excess
part vanishing at zero voltage, $G(x)=G^\mathrm{eq}(x)+
G^\mathrm{ex}(x)$,
\begin{eqnarray}
      &&G^\mathrm{eq} (x)
      =[\sin(k_{\rm\scriptscriptstyle F}x)]/\pi x,
      \label{1p_corr_eq}
      \\
      &&G^\mathrm{ex} (x) =
      e^{-i(k_{\rm\scriptscriptstyle F}+
      k_{\scriptscriptstyle V}/2)x} \,
      [\sin(k_{\scriptscriptstyle V}x/2)]/\pi x,
      \label{1p_corr_ex}
\end{eqnarray}
with $k_{\scriptscriptstyle V} = k_{\rm\scriptscriptstyle F}
(eV/2\varepsilon_{\rm \scriptscriptstyle F})$. The equilibrium
part of the pair correlator $g_{\vec\sigma}^\mathrm{eq}(x,y) =
\langle \sigma_1|\sigma_4\rangle\langle\sigma_2|\sigma_3\rangle
[G^\mathrm{eq}(0)]^2 -\langle\sigma_1|\sigma_3\rangle
\langle\sigma_2|\sigma_4\rangle |G^\mathrm{eq}(x-y)|^2$
describes the exchange correlations between two fermions with
singlet correlations decaying on the Fermi wave length
$\lambda_{\rm\scriptscriptstyle F}$. The excess part
\begin{eqnarray}
   &&g_{\vec\sigma}^\mathrm{ex}(x,y) = \langle\sigma_1|\sigma_4\rangle
   \langle\sigma_2|\sigma_3\rangle[G^\mathrm{ex}(0)]^2
   \nonumber \\
   &&\qquad\qquad
   -\,\langle\sigma_1|\sigma_3\rangle \langle\sigma_2|\sigma_4\rangle
   |G^\mathrm{ex}(x-y)|^2
   \label{g_ex_s}
\end{eqnarray}
associated with the additional injected electrons exhibits
singlet correlations on the much larger scale
$\lambda_{\scriptscriptstyle V}=\lambda_{\rm
\scriptscriptstyle F}(\varepsilon_{\rm \scriptscriptstyle F}
/eV)$ (an additional mixed term describing the deformation
of the equilibrium exchange hole \cite{madelung} due to the
bias is not relevant to our discussion and we ignore it here).
We conclude, that the excess particles injected by the
reservoir form a {\it stream of singlet-correlated pairs}.
Furthermore, again due to the Fermi statistics,
these singlet-pairs propagate as a regular sequence of wave
packets separated by the single particle correlation time
\cite{lesoviklevitov} $\tau_V=\hbar/eV$ corresponding to the
singlet correlation length $\lambda_{\scriptscriptstyle V}$. The
singlet-pairs are conveniently described through the state
$|\Psi_\mathrm{in}\rangle =|e_{\uparrow\downarrow}^\mathrm{sg}
\rangle_\mathrm{s}$ (with index `s' for `source') or the
corresponding wave function $\Psi_\mathrm{in} =
\phi_\mathrm{s}^{\scriptscriptstyle 1}
\phi_\mathrm{s}^{\scriptscriptstyle 2}
\chi_\mathrm{sg}^{\scriptscriptstyle 12}$, with the spin-singlet
wave function $\chi_\mathrm{sg}^{\scriptscriptstyle 12}
=[\chi_\uparrow^{\scriptscriptstyle 1}
\chi_\downarrow^{\scriptscriptstyle 2}
-\chi_\downarrow^{\scriptscriptstyle 1}
\chi_\uparrow^{\scriptscriptstyle 2}]/\sqrt{2}$ and upper indices
1 and 2 denoting the particle number. Note that the emission of
singlet-pairs from the normal reservoir follows naturally from the
identical orbital wave function describing electrons with opposite
spins within the reservoir.

Next, we analyse the scattered state propagating in the
two leads of the fork in Fig.\ \ref{fig:fork}.
The pair correlation function describing particles
propagating in different leads ($x$ in `u' and $y$ in `d')
takes the form
\begin{eqnarray}
   &&
   g_{\vec\sigma}(x,y)=
   \langle\sigma_1|\sigma_4\rangle\langle\sigma_2|\sigma_3\rangle
   \,G_\mathrm{u}(0)G_\mathrm{d}(0)
   \nonumber\\
   &&\qquad\quad\>\>\>-\,
   \langle\sigma_1|\sigma_3\rangle\langle\sigma_2|\sigma_4\rangle
   |G_\mathrm{ud}(x-y)|^2,
   \label{pair_f}
\end{eqnarray}
and has to be calculated with the scattering states
\begin{eqnarray}
   && \hat\Psi_\mathrm{u}=\int\frac{d\epsilon}{\sqrt{hv_\epsilon}}
   \bigl[\bigl(t_\mathrm{su} \hat{c}_{\epsilon}
   +r_\mathrm{u} \hat{a}_{\epsilon}
   +t_\mathrm{du}\hat{b}_{\epsilon} \bigr)e^{ikx}
   +\hat{a}_{\epsilon} e^{-ikx} \bigr],
   \nonumber \\
   && \hat \Psi_\mathrm{d}=\int\frac{d\epsilon}{\sqrt{hv_\epsilon}}
   \bigl[\bigl(t_\mathrm{sd}\hat{c}_{\epsilon}
   +t_\mathrm{ud}\hat{a}_{\epsilon}
   +r_\mathrm{d}\hat{b}_{\epsilon}\bigr)e^{ikx}
   +\hat{b}_{\epsilon}e^{-ikx}\bigr],
   \nonumber
\end{eqnarray}
where $\hat{a}_\epsilon$, $\hat{b}_\epsilon$, and
$\hat{c}_\epsilon$ denote the annihilation operators for spinless
electrons at energy $\epsilon$ in leads `u', `d', and `s' and with
a time evolution $\propto \exp(-i\epsilon t/\hbar)$ (here,
$t_\mathrm{su}$ ($t_\mathrm{du}$) and $t_\mathrm{sd}$
($t_\mathrm{ud}$) describe particle transmission from the source
(down) lead into the `up' lead and from the source (up) lead into
the `down' lead; $r_\mathrm{u}$, $r_\mathrm{d}$ denote the
reflection amplitudes into leads `u' and `d'). The orbital
one-particle correlators $G_\mathrm{x}(x)=\langle\hat
\Psi_\mathrm{x}^\dagger(x)\hat\Psi_\mathrm{x}(0)\rangle =
G^\mathrm{eq}(x) + T_\mathrm{x} G^\mathrm{ex}(x)$ describe
particles with coordinate $x$ residing in the same lead `x' equal
to `u' or `d' ($T_\mathrm{u} = |t_\mathrm{su}|^2$ and
$T_\mathrm{d} = |t_\mathrm{sd}|^2$ describe the transmission
probabilities into the `up' and `down' lead). The one-particle
cross correlator between leads `u' and `d' takes the form
$G_\mathrm{ud}(x-y)=\langle\hat \Psi_\mathrm{u}^\dagger(x)
\hat\Psi_\mathrm{d}(y)\rangle = t_\mathrm{su}^\ast t_\mathrm{sd}
G^\mathrm{ex}(x-y)$ with coordinates $x$ and $y$ residing in the
leads `u' and `d'. The excess part of the pair correlation
function reads
\begin{eqnarray}
   &&g_{\vec\sigma}^\mathrm{ex}(x,y)
   =T_\mathrm{u}T_\mathrm{d}\bigl[
   \langle\sigma_1|\sigma_4\rangle \langle\sigma_2|\sigma_3\rangle
   [G^\mathrm{ex}(0)]^2
   \nonumber\\
   &&\qquad\qquad-\,\langle\sigma_1|\sigma_3\rangle
   \langle\sigma_2|\sigma_4\rangle
   |G^\mathrm{ex}(x-y)|^2\bigr].
   \label{pair_res}
\end{eqnarray}
This result is identical in form with the excess
pair correlator (\ref{g_ex_s}) in the source lead, however,
it now describes the correlation between a singlet-pair
split into the leads `u' and `d'. Note the preservation
of the singlet-correlations which are maximal for $x= y$ and
decay on a distance $|x-y| \sim \lambda_{\scriptscriptstyle V}$,
where the coordinates $x$ and $y$ belong to different leads.
The scattering state describing the propagation of the singlet-pair
behind the splitter can be written in the form
\begin{eqnarray}
   &&|\Psi_\mathrm{out}^\mathrm{ee}\rangle=t_\mathrm{su}^2
   |e_{\uparrow\downarrow}^\mathrm{sg}\rangle_\mathrm{u}
   |0\rangle_\mathrm{d}+t_\mathrm{sd}^2
   |0\rangle_\mathrm{u}
   |e_{\uparrow\downarrow}^\mathrm{sg}\rangle_\mathrm{d}
   \label{ee_scat}\\
   &&\qquad\quad+\sqrt{2} t_\mathrm{su}t_\mathrm{sd}
   \bigl[|e_\uparrow\rangle_\mathrm{u}
   |e_\downarrow\rangle_\mathrm{d}-
   |e_\downarrow\rangle_\mathrm{u}|e_\uparrow\rangle_\mathrm{d}
   \bigr],
   \nonumber
\end{eqnarray}
where the first two terms describe the propagation of the
singlet-pair $|e_{\uparrow\downarrow}^\mathrm{sg}\rangle_\mathrm{x}$
in leads `x' equal `u' or `d', with a wave function of the form
$\phi_\mathrm{x}^{\scriptscriptstyle 1}
\phi_\mathrm{x}^{\scriptscriptstyle 2}
\chi_\mathrm{sg}^{\scriptscriptstyle 12}$.
The last term describes a singlet-pair split between the
`up' and `down' leads with a wave function
$[\phi^{\scriptscriptstyle 1}_\mathrm{u}
\phi^{\scriptscriptstyle 2}_\mathrm{d}+\phi^{\scriptscriptstyle
1}_\mathrm{d}\phi^{\scriptscriptstyle 2}_\mathrm{u}]
\chi^{\scriptscriptstyle 12}_\mathrm{sg}$.
A coincidence measurement of electrons in leads `u' and `d'
projects the scattered state $|\Psi_\mathrm{out}^\mathrm{ee}\rangle$
onto this spin-entangled component with spatially separated
electrons in leads `u' and~`d'.

In the tunneling limit \cite{beenakker_03} ($T_\mathrm{u}
\sim 1$ and $T_\mathrm{d}\ll 1$) most of the incoming
singlet pairs propagate into the well conducting `up'
lead and only rarely (with amplitude $t_\mathrm{su}
t_\mathrm{sd}$) split into both leads. The absence of
an electron in the `up' lead then manifests itself
as the presence of a hole and it is favorable to
go over to a hole representation,
\begin{eqnarray}
  &&|\Psi_\mathrm{out}^\mathrm{he}\rangle
  = t_\mathrm{su}^2
  |0\rangle_\mathrm{u}|0\rangle_\mathrm{d}
  +t_\mathrm{sd}^2
  |h_{\uparrow\downarrow}^\mathrm{sg}\rangle_\mathrm{u}
  |e_{\uparrow\downarrow}^\mathrm{sg}\rangle_\mathrm{d}
  \label{eh_scat}\\
   &&\qquad\quad +\sqrt{2}t_\mathrm{su}t_\mathrm{sd}
  \bigl[ |h_\downarrow\rangle_{\mathrm{u}}
  |e_\downarrow\rangle_\mathrm{d}
  -|h_\uparrow\rangle_\mathrm{u}
  |e_\uparrow\rangle_\mathrm{d}\bigr]
  \nonumber
\end{eqnarray}
and the hole current $\hat I^\mathrm{h}_\mathrm{u} =(2e^2/h)V- \hat
I_\mathrm{u}^\mathrm{e}$. The first term in (\ref{eh_scat}) (the
component $\alpha$ in Fig.\ 1(b)) describes a filled Fermi see in
the upper lead combined with a vacuum state in the lower lead, and
hence no particle can be detected. The second term (component
$\delta$) accounts for the  rare processes where both electrons
propagate to the `down' lead; its contribution spoils the
maximum violation of the BI and restricts the use of the
tunneling limit. The most relevant terms are the last two
($\beta$ and $\gamma$) describing the splitting of the
singlet-pair between the two leads and the formation of a
spatially separated spin-entangled electron-hole pair with the
hole (electron) propagating in the upper (lower) lead;
this electron-hole component is detectable in a coincidence
measurement using a hole (particle) detector in the
upper (lower) lead.

\section{Bell inequalities}

\subsection{Bell inequality with current correlators}

The original goal in setting up the Bell inequalities \cite{bell}
was to devise a scheme allowing for the differentiation
between classical correlations appearing in a local hidden
variable theory and non-local correlations as they show up
within a quantum mechanical framework. Accordingly, early
experiments in optics addressed these fundamental questions
dealing with the validity of quantum mechanics. Recently,
Bell inequalities have been discussed in the context of
mesoscopic electronic devices. One should admit, that
the corresponding electronic setups are probably less
suitable for addressing fundamental issues of quantum
mechanics. In a more pragmatic approach, the Bell
inequalities in mesoscopic physics are used as
a test for the presence of entanglement or even for a
quantitative measurement of the degree of entanglement
between quantum degrees of freedom.

Defining appropriate Bell inequalities in mesoscopic physics
is nevertheless a non-trivial issue as those observables
suitable for direct measurement in optics are not necessarily
available in mesoscopics; this is why we attempt an extended
and detailed analysis below. In this context, we keep the
discussion on a level where fundamental and practical issues
can be easily identified.

The explicit form of the Bell inequality we are going to use
in the present paper has been introduced by Clauser and Horne
\cite{clauserhorne} based on the original discussion of Bell
\cite{bell}. It derives from the lemma saying that, given a
set of real numbers $x,\bar{x},y,\bar{y},X,Y$ with $|x/X|$,
$|\bar{x}/X|$, $|y/Y|$, and $|\bar{y}/Y|$ restricted to
the interval $[0,1]$, the inequality
\begin{equation}
   |xy-x\bar{y}+\bar{x}y+\bar{x}\bar{y}|
   \leq 2 |XY|
   \label{lemma}
\end{equation}
holds true.
In the Bell type setup of Fig.\ \ref{fig:fork}(a) one measures
the spin-projected electronic currents ${I}^\mathrm{e}_i$,
$i=1,\dots,4$, and defines the quantities $x = {I}_1^\mathrm{e} -
{I}_3^\mathrm{e}$, $X= {I}_1^\mathrm{e} +{I}_3^\mathrm{e}$
and $y = {I}_2^\mathrm{e}-{I}_4^\mathrm{e}$, $Y={I}_2^\mathrm{e}
+{I}_4^\mathrm{e}$ for fixed orientations ${\bf a}$ and ${\bf b}$
of the polarizers (and similarly $\bar{x}$ and $\bar{y}$ for
orientations $\bar{\bf a}$ and $\bar{\bf b}$).
Our Bell setup then measures the correlations
\begin{eqnarray}
   {\cal C}^\mathrm{ee}_{ij}({\bf a},{\bf b};\tau)
   &=&\langle {I}_i^\mathrm{e} (\tau)
            {I}_j^\mathrm{e} (0) \rangle_\lambda
   \label{CC}\\
   &\equiv& \int d\lambda \, \rho(\lambda) \,
   {I}_i^\mathrm{e}({\bf a},\lambda;\tau)
   {I}_j^\mathrm{e} ({\bf b}, \lambda;0) \nonumber
\end{eqnarray}
between the spin-currents $I^\mathrm{e}_i$, $i=1,3$,
in lead `u' projected onto the directions $\pm{\bf a}$
and their partners $I^\mathrm{e}_j$, $j=2,4$, in lead `d'
projected onto $\pm{\bf b}$. Within a local hidden variable
($\lambda$) theory, the average $\langle\dots\rangle_\lambda$
is taken with respect to the distribution function $\rho(\lambda)$;
specifying a theoretical framework such as quantum mechanics,
these averages are replaced by quantum mechanical averages.
Using the above definitions of $x$, $y$, $X$, and $Y$, we
obtain the current difference correlator
\begin{eqnarray}
   &&E({\bf a},{\bf b};\tau)
   =\frac{\langle[{I}^\mathrm{e}_1(\tau)
   -{I}^\mathrm{e}_3(\tau)]
   [{I}^\mathrm{e}_2(0)
   -{I}^\mathrm{e}_4(0)]\rangle_\lambda}
   {\langle[{I}^\mathrm{e}_1(\tau)
   +{I}^\mathrm{e}_3(\tau)]
   [{I}^\mathrm{e}_2(0)
   +{I}^\mathrm{e}_4(0)]\rangle_\lambda},
   \nonumber\\
   && \qquad\qquad\quad
   = \frac{{\cal C}^\mathrm{ee}_{12}
   -{\cal C}^\mathrm{ee}_{14}
   -{\cal C}^\mathrm{ee}_{32}
   +{\cal C}^\mathrm{ee}_{34}}
   {{\cal C}^\mathrm{ee}_{12}
   +{\cal C}^\mathrm{ee}_{14}
   +{\cal C}^\mathrm{ee}_{32}
   +{\cal C}^\mathrm{ee}_{34}},
   \label{E0}
\end{eqnarray}
and evaluating these for different combinations of directions
${\bf a}$, $\bar{\bf a}$, ${\bf b}$, and $\bar{\bf b}$ we can
combine them into the Bell inequality
\begin{equation}
   E_{\rm\scriptscriptstyle BI}(\tau)
   \!=\! |E({\bf a},{\bf b})-E({\bf a},\bar{\bf b})
   +E(\bar{\bf a},{\bf b})+E(\bar{\bf a},\bar{\bf b})| \leq 2.
   \label{BI}
\end{equation}

We can further process the difference correlator (\ref{E0}) and
separate the current correlators into an irreducible part
$C^\mathrm{ee}_{ij} ({\bf a},{\bf b};\tau) = \langle \delta
{I}_i^\mathrm{e}(\tau) \, \delta {I}_j^\mathrm{e}(0)
\rangle_\lambda$ with $\delta {I}_i^\mathrm{e}(\tau) =
{I}_i^\mathrm{e}(\tau) - \langle {I}_i^\mathrm{e}
\rangle_\lambda$ and a product of average currents and rewrite
$E({\bf a},{\bf b};\tau)$ in the form
\begin{equation}
   E({\bf a},{\bf b};\tau) =
   \frac{C^\mathrm{ee}_{12}
   -C^\mathrm{ee}_{14}
   -C^\mathrm{ee}_{32}
   +C^\mathrm{ee}_{34}
   +\Lambda^\mathrm{ee}_-}
   {C^\mathrm{ee}_{12}
   +C^\mathrm{ee}_{14}
   +C^\mathrm{ee}_{32}
   +C^\mathrm{ee}_{34}
   +\Lambda^\mathrm{ee}_+},
   \label{E1}
\end{equation}
with $\Lambda^\mathrm{ee}_\pm = [\langle{I}^\mathrm{e}_1
\rangle_\lambda \pm\langle{I}^\mathrm{e}_3\rangle_\lambda]
[\langle{I}^\mathrm{e}_2\rangle_\lambda\pm\langle
{I}^\mathrm{e}_4\rangle_\lambda]$. The average currents are
related via $\langle {I}^\mathrm{e}_1\rangle_\lambda= \langle
{I}^\mathrm{e}_3\rangle_\lambda =\langle
{I}^\mathrm{e}_\mathrm{u}\rangle_\lambda/2$ and $\langle
{I}^\mathrm{e}_2\rangle_\lambda=\langle {I}^\mathrm{e}_4\rangle_\lambda
=\langle {I}^\mathrm{e}_\mathrm{d}\rangle_\lambda/2$ and thus
$\Lambda^\mathrm{ee}_-=0$, $\Lambda^\mathrm{ee}_+=\langle
{I}^\mathrm{e}_\mathrm{u}\rangle_\lambda \langle
{I}^\mathrm{e}_\mathrm{d}\rangle_\lambda$,
\begin{equation}
   E({\bf a},{\bf b};\tau) =
   \frac{C^\mathrm{ee}_{12}
   -C^\mathrm{ee}_{14}
   -C^\mathrm{ee}_{32}
   +C^\mathrm{ee}_{34}}
   {C^\mathrm{ee}_{12}
   +C^\mathrm{ee}_{14}
   +C^\mathrm{ee}_{32}
   +C^\mathrm{ee}_{34}
   +\langle
   {I}^\mathrm{e}_\mathrm{u}\rangle_\lambda \langle
   {I}^\mathrm{e}_\mathrm{d}\rangle_\lambda}.
   \label{E}
\end{equation}
In the tunneling limit, the electronic currents
${I}^\mathrm{e}_{i}$ in the `up' lead are replaced by hole
currents ${I}^\mathrm{h}_{i} = I_\mathrm{max}^\mathrm{e}
-{I}^\mathrm{e}_{i}$, where $I_\mathrm{max}^\mathrm{e}$ denotes
the current in the open channel (within a quantum mechanical
framework, $I_\mathrm{max}^\mathrm{e}= (e^2/h)V$).

\subsection{Bell inequality with coincidence probabilities}

The Bell inequality (\ref{BI}) explicitely depends on the delay
time $\tau$ which appears naturally through the time dependence
in the current correlators in (\ref{E0}) and (\ref{E}). This
implies that the violation of the Bell inequality depends on
the delay- or measurement time of the experiment, a feature not
encountered in traditional optical setups. In optics, the quantity
usually involved in this type of analysis is the coincidence
probability $P_{ij}({\bf a},{\bf b})$ for the simultaneous
detection of two photons with polarizations along ${\bf a}$
and ${\bf b}$.
The indices $i$ and $j$ specify the directions along and
perpendicular to the polarizer --- here, these correspond
to the spin-up and spin-down events triggering a signal in
the detectors $i=1,3$ or $j=2,4$. In optics, these coincidence
probabilities are directly measurable and can be combined
into a Bell inequality $\tilde{E}_{\rm\scriptscriptstyle
BI}=|\tilde{E}({\bf a},{\bf b})-\tilde{E}({\bf a},\bar{\bf b})
+\tilde{E}(\bar{\bf a},{\bf b})+\tilde{E}(\bar{\bf a},\bar{\bf b})|
\leq 2$ with
\[
   \tilde{E}({\bf a},{\bf b}) =
    P_{12}({\bf a},{\bf b})
   -P_{14}({\bf a},{\bf b})
   -P_{32}({\bf a},{\bf b})
   + P_{34}({\bf a},{\bf b}).
\]
Contrary to the situation in optics (where photons are
annihilated in the detector), the coincidence probability
$P_{ij}({\bf a},{\bf b})$ is not a directly measurable
quantity in mesoscopics, where the observables of choice
are the (electron or hole) currents $I^\mathrm{e,h}_i$.
The expression of $P_{ij}({\bf a},{\bf b})$ through
measurable currents then requires some care and we
provide a detailed discussion here in order to compare
the approach based on coincidence probabilities with
the one based on current correlators, see section III.A.

A natural way to define a coincidence probability in
mesoscopics is through particle number correlators
\begin{equation}
   {\cal K}_{ij}^\mathrm{ee}({\bf a},{\bf b};\tau)
   =\langle {N}_i^\mathrm{e}(\tau)
   {N}_j^\mathrm{e}(\tau)\rangle_\lambda
   \label{CNN}
\end{equation}
where
\begin{equation}
   {N}_i^\mathrm{e}(\tau)
   = \int_0^\tau dt \, {I}_i^\mathrm{e}(t)
   \label{N}
\end{equation}
is the number of electrons counted by the detector $i$
during the accumulation time $\tau$. Here, we are
interested in the simultaneous detection of two
particles, one appearing in the upper lead `u',
the other in the down lead `d'.  In order to
obtain such a coincidence probability from the
number correlator (\ref{CNN}) we have to restrict the
accumulation time such that only events $\{0,0\}$
(no particles) $\{1,0\}$ (one particle in detector $i$),
$\{0,1\}$ (one particle in detector $j$), and $\{1,1\}$
(two particles, one in detector $i$ and one in $j$) are
accounted for; out of these, the coincident events
$\{1,1\}$ then contribute to ${\cal K}_{ij}^\mathrm{ee}$.
Events of the type $\{2,1\}$, $\{1,2\}$, $\dots$
have to be avoided through proper time limitation.
Restricting the accumulation time $\tau$ to a value
$\tau_1$ such that at most one particle is counted,
$\langle {N}_i^\mathrm{e}(\tau_1)\rangle_\lambda \leq 1$,
and using a proper normalization, we can find an
expression for the coincidence probability in terms
of the number correlator (\ref{CNN}),
\begin{equation}
   P_{ij}^\mathrm{ee}({\bf a},{\bf b}; \tau_1)
   \equiv \frac{{\cal K}_{ij}^\mathrm{ee}}
   {{\cal K}_{12}^\mathrm{ee}
   +{\cal K}_{14}^\mathrm{ee}
   +{\cal K}_{32}^\mathrm{ee}
   +{\cal K}_{34}^\mathrm{ee}}
   \label{P}
\end{equation}
with all correlators ${\cal K}_{ij}^\mathrm{ee}({\bf a},{\bf b},
\tau_1)$ evaluated at a fixed time $\tau_1$ and fixed
directions ${\bf a}$ and ${\bf b}$, cf.\ (\ref{CNN}).
The condition $\langle {N}_i^\mathrm{e}(\tau_1)\rangle \leq 1$
requires that $\tau_1$ is smaller than the time $\tau_I
= e/\max_i[\langle I^\mathrm{e}_i\rangle_\lambda]$
between subsequent events, $\tau_1 \ll \tau_I$. Note, that
during the short accumulation time $\tau_1$ {\it less} than one
particle contributes on average, however, this is corrected
for by the proper normalization. Also, we note that our
definition (\ref{P}) for the coincidence rate is not
identical to the one introduced in Ref.\ \onlinecite{samuelsson_03}
and used in Refs.\ \onlinecite{beenakker_03,ssb_04};
we will return to this point later.

The quantity (\ref{P}) can be further processed in the limit
of rare events, i.e., in the tunneling limit ($T_\mathrm{d}\ll
1$). In this situation it is advantageous to go over to
electron-hole correlators \cite{beenakker_03} with the
hole current defined via ${I}^\mathrm{h}_i =
I_\mathrm{max}^\mathrm{e}-{I}^\mathrm{e}_i$ (note that in Ref.\
\onlinecite{chtchelkatchev_02} the electronic currents
in the two normal leads originate from the low-rate
emission of Cooper pairs from the superconductor and
hence are {\it both small}; alternatively, one may
view these excess currents as arising from
Andreev-scattering at the normal-superconductor
interface where an incident electron is reflected
as a hole). Again, the electron-hole number correlators
entering (\ref{P}) are split into an irreducible
and a reducible part,
\begin{equation}
   {\cal K}_{ij}^\mathrm{he}(\tau_1)
   = K_{ij}^\mathrm{he}(\tau)
   +\langle {I}_i^\mathrm{h}\rangle_\lambda
   \langle {I}_i^\mathrm{e}\rangle_\lambda \tau_1^2.
   \label{CNNdec}
\end{equation}
with the irreducible number correlator defined as
\begin{equation}
   K_{ij}^\mathrm{he}(\tau)\equiv\langle
   \delta  N_i^\mathrm{h}(\tau)
   \delta  N_j^\mathrm{e}(\tau)\rangle_\lambda
   =\!\int_0^\tau\!\!\!\! dt \!\int_{-t}^t \!\!\!\!
   dt^\prime \, C_{ij}^\mathrm{he}(t^\prime)
   \label{Kdef}
\end{equation}
Next, we can make use of the fact that the irreducible
current correlator $C_{ij}^\mathrm{he} (t')$ rapidly
decays on a time scale $\tau_c$ (within a quantum
mechanical framework, $\tau_c = \tau_V = \hbar/eV$
is the single particle correlation time).  For $\tau_1
\gg \tau_c$, we can approximate the irreducible
part of ${\cal K}_{ij}^\mathrm{he} (\tau_1)$ by the
time-independent zero-frequency noise correlator
$S_{ij}^\mathrm{he}(\omega=0)$ (cf.\ Ref.\
\onlinecite{chtchelkatchev_02}),
\begin{equation}
   {\cal K}_{ij}^\mathrm{he}(\tau_1)
   =S_{ij}^\mathrm{he}(\omega=0) \tau_1
   +\langle {I}_i^\mathrm{h}\rangle_\lambda
   \langle {I}_i^\mathrm{e}\rangle_\lambda \tau_1^2.
   \label{CNNzfn}
\end{equation}
Inserting this expression into the correlators $\tilde{E}$
we arrive at the expression analogous to the one introduced
in Ref.~\onlinecite{chtchelkatchev_02}
\begin{equation}
   \tilde{E}({\bf a},{\bf b};\tau_1) =
   \frac{S^\mathrm{he}_{12}
   -S^\mathrm{he}_{14}
   -S^\mathrm{he}_{32}
   +S^\mathrm{he}_{34}}
   {S^\mathrm{he}_{12}
   +S^\mathrm{he}_{14}
   +S^\mathrm{he}_{32}
   +S^\mathrm{he}_{34} +\Lambda_+^\mathrm{he} \tau_1}.
   \label{bESL}
\end{equation}
In a final step, we demonstrate that we can ignore
the current product term $\Lambda_+^\mathrm{he} \tau_1$
in the tunneling limit. In order to gain more insight,
we analyze the situation theoretically within a quantum
mechanical framework for non-interacting electrons:
with $S_{ij}^\mathrm{he}(\omega=0) = |\langle {\bf a}_i
|{\bf b}_j\rangle|^2 (e^2/h) T_\mathrm{u} T_\mathrm{d}
\,eV$ and $\langle {I}_i^\mathrm{h}\rangle \approx
\langle {I}_j^\mathrm{e}\rangle = (e^2/h) T_\mathrm{d}\,
V$ (we assume that $1-T_\mathrm{u} \approx T_\mathrm{d}$;
the vectors ${\bf a}_i$ and ${\bf b}_j$ denote the
directions associated with the detectors $i$ and
$j$) we find that ${\cal K}_{ij}^\mathrm{he}(\tau_1)
= (e^2/h) \tau_1 T_\mathrm{d} \, eV (|\langle {\bf a}_i
|{\bf b}_j\rangle|^2T_\mathrm{u}+\tau_1 T_\mathrm{d}
/2\pi\tau_V)$. The second term is due to the current
product term and can be dropped provided that $\tau_1
T_\mathrm{d}/2\pi\tau_V \ll 1$ (the appearance of the
small factor $T_\mathrm{d}$ is due to the use of
hole currents). In the tunneling limit,
$\tau_V/T_\mathrm{d} = \tau_I \gg \tau_V$ and we find
a large time window $\tau_c = \tau_V \ll \tau_1 \ll \tau_I$
within which we can choose our accumulation time
$\tau_1$ such that both the definition (\ref{P}) and
the separation (\ref{CNNzfn}) can be properly installed
and the current product term is small. It is crucial
to understand that under these circumstances the
accumulation time $\tau_1$ appears as a {\it theoretical}
quantity which is needed only in the derivation of the
Bell parameter; once we have demonstrated that $\tau_1$
can be chosen such that the current product term is small,
the latter can be dropped and $\tau_1$ {\it disappears}
from the Bell parameter.

While the above idealized theoretical consideration serves
as a guideline, the situation in a real experiment may
be complicated due to interactions and other effects
which change the value of the correlation time $\tau_c$,
i.e., $\tau_c$ may differ substantially from $\tau_V$.
Therefore, in an actual experiment testing the Bell
inequalities the decay time $\tau_c$ of the correlator
should be measured independently in order to verify that
the second term in (\ref{CNNzfn}) involving the product
of average currents is indeed small on this time scale
and thus can be dropped. Fortunately, in the tunneling
limit the admissible time window $\tau_c \ll \tau_1 \ll
\tau_I$ is large and this test can be carried out
with only an approximate knowledge of the correlation
time $\tau_c$.

Once we have verified that we can drop the correction
term $\langle {I}_i^\mathrm{h} \rangle_\lambda \langle
{I}_i^\mathrm{e}\rangle_\lambda \tau_1^2$ in (\ref{CNNzfn})
we can express the coincidence probabilities through the
(time-independent) zero-frequency noise correlators
$S_{ij}^\mathrm{he} (\omega=0)$ alone,
\begin{equation}
   P_{ij}^\mathrm{he}({\bf a},\!{\bf b})
   =\frac{S_{ij}^\mathrm{he}}
   {S_{12}^\mathrm{he}
   +S_{14}^\mathrm{he}
   +S_{32}^\mathrm{he}
   +S_{34}^\mathrm{he}};
   \label{PS}
\end{equation}
the correlator $\tilde{E}({\bf a},{\bf b})$ then takes the
simple form
\begin{equation}
   \tilde{E}({\bf a},{\bf b}) =
   \frac{S^\mathrm{he}_{12}
   -S^\mathrm{he}_{14}
   -S^\mathrm{he}_{32}
   +S^\mathrm{he}_{34}}
   {S^\mathrm{he}_{12}
   +S^\mathrm{he}_{14}
   +S^\mathrm{he}_{32}
   +S^\mathrm{he}_{34}}.
   \label{bES}
\end{equation}
The results (\ref{PS}) and (\ref{bES}) now are
independent of time, although the original expression
(\ref{P}) involved the time restriction $\tau_1 \ll
\tau_I$; their use is restricted to the tunneling
limit as only in this case the correction term $\propto
\Lambda^\mathrm{he}_+$ in the denominator of $\tilde{E}$
can be dropped.

Let us next concentrate on the general situation away from the
tunneling limit which turns out quite different. In this
case, there is no advantage in going over to hole
currents and we work with electronic currents in
both leads. Second, in the absence of a small
tunneling probability we have no separation of time
scales and $\tau_c \sim \tau_I$ are of the same order,
hence the accumulation time $\tau_1$
has to be smaller than (or at most of the order of) the
single particle correlation time $\tau_c$, $\tau_1 \leq
\tau_c$. As a consequence, one {\it cannot} express the
coincidence probability $P_{ij}^\mathrm{ee}$ through the
zero-frequency noise correlator $S_{ij}^\mathrm{ee}$.
Instead, we make use of the irreducible number correlator
$K_{ij}^\mathrm{ee}(\tau)$, cf.\ (\ref{Kdef}),
and rewrite the coincidence probability in the form
\begin{equation}
   P_{ij}^\mathrm{ee}({\bf a},{\bf b};\tau_1)
   =\frac{K_{ij}^\mathrm{ee}+\langle I_i^\mathrm{e}
   \rangle_\lambda \langle I_j^\mathrm{e}
   \rangle_\lambda \tau_1^2 }
   {K_{12}^\mathrm{ee}
   +K_{14}^\mathrm{ee}
   +K_{32}^\mathrm{ee}
   +K_{34}^\mathrm{ee} + \Lambda^\mathrm{ee}_+ \tau_1^2}.
   \label{PK}
\end{equation}
Combining the coincidence probabilities into the expression
$\tilde{E}({\bf a},{\bf b})$, the products of average
currents cancel in the numerator, $\Lambda^\mathrm{ee}_-=0$,
however, these products {\it do not} cancel in the denominator
and restrict the violation of the Bell inequality to
small time intervals (see below). The final correlator entering
the Bell parameter $\tilde{E}_{\rm\scriptscriptstyle BI}$ then
takes the form
\begin{equation}
   \tilde{E}({\bf a},{\bf b};\tau_1) =
   \frac{K^\mathrm{ee}_{12}
   -K^\mathrm{ee}_{14}
   -K^\mathrm{ee}_{32}
   +K^\mathrm{ee}_{34}}
   {K^\mathrm{ee}_{12}
   \!+\!K^\mathrm{ee}_{14}
   \!+\!K^\mathrm{ee}_{32}
   \!+\!K^\mathrm{ee}_{34}
   \!+\!\langle I_\mathrm{u}^\mathrm{e}
   \rangle_\lambda \langle I_\mathrm{d}^\mathrm{e}
   \rangle_\lambda \tau_1^2}.
   \label{bE}
\end{equation}
Comparing (\ref{E}) with (\ref{bE}) it turns out that both
Bell parameters $E_{\rm\scriptscriptstyle BI}$ and
$\tilde{E}_{\rm\scriptscriptstyle BI}$, when derived carefully,
are based on the same correlations, once expressed
directly through currents, the other time through
number correlators. In particular, both Bell
parameters depend on time and, as we will see below, the
violation of the Bell inequalities is limited to times
$\tau_1 < \tau_c \sim \tau_I$. Quite interestingly, this
coincides with the time restriction imposed by our definition
(\ref{P}) of the coincidence probability in terms of
number correlators --- it turns out that the Bell
inequality can be violated during those times $\tau_1$
where the (normalized) number correlator
assumes the meaning of a coincidence probability.

The correlator (\ref{bE}) has first been introduced
and used in Ref.\ \onlinecite{chtchelkatchev_02}
in order to analyze the entanglement of electrons
injected by a superconductor into a normal metal
fork. On the other hand, the simplified version
(\ref{bES}) valid in the tunneling limit and first
introduced (via an alternative route) in
Ref.\ \onlinecite{samuelsson_03} has been
successfully applied in Ref.\ \onlinecite{beenakker_03}.
However, its application in Ref.\ \onlinecite{beenakker_03b}
to a situation away from the tunneling regime led
to the appearance of a `spurious amplification
factor' in the relation between the Bell parameter
and the concurrence, a consequence of ignoring
the presence of the current product term $\propto
\langle \hat{I}_\mathrm{u}^\mathrm{e} \rangle
\langle \hat{I}_\mathrm{d}^\mathrm{e} \rangle$.
Instead, the subsequent use \cite{beenakker_03b} of
the full expression (\ref{bE}) removed this problem
and established the violation of electronic Bell
inequalities in the non-tunneling regime at short
times $\tau < \tau_V = \hbar/eV$.

In the discussion above, we have been careful to separate
theoretical from experimental input in the construction
of the Bell parameter and have provided
final expressions involving only experimental input.
Depending on the chosen variables and on the physical
situation, we have seen that a short time measurement
is required in general; in the tunneling limit, an
approximate determination of the correlation time
$\tau_c$ together with the zero-frequency noise
correlator is sufficient. It is interesting to compare
our point of view expressed in the above derivation
with the approach introduced in Ref.\
\onlinecite{samuelsson_03}, where the authors derive
a Bell parameter which does not involve a short time
measurement. While this scheme works the same way as
ours in the tunneling limit (rough estimate of $\tau_c$
and knowledge of the zero-frequency noise are sufficient),
it does require the additional {\it precise} knowledge of the
correlation time $\tau_c$ away from the tunneling limit,
which either requires an accurate theoretical evaluation
for the device at hand or again necessitates a short
time measurement.

In Refs.\ \onlinecite{samuelsson_03,ssb_04}
the coincidence probability $P_{ij}({\bf a},{\bf b})$
has been defined as an equal time correlator of the
form,
\begin{eqnarray}
   P_{ij}({\bf a}, {\bf b}) &\propto&
   \langle \hat{a}_i^\dagger(t) \hat{b}_j^\dagger(t)
   \hat{b}_j(t) \hat{a}_i(t)\rangle \label{P_But} \\
   &\propto& \langle \hat{I}^\mathrm{e}_i(t)
   \hat{I}^\mathrm{e}_j(t) \rangle =
   \langle \hat{I}^\mathrm{e}_i\rangle
   \langle\hat{I}^\mathrm{e}_j \rangle
   +\langle \delta \hat{I}^\mathrm{e}_i(t)
   \delta \hat{I}^\mathrm{e}_j(t) \rangle.
   \nonumber
\end{eqnarray}
Here, the operators $\hat{a}_i$ and $\hat{b}_j$
annihilate electrons in the upper and lower leads
in the detectors $i=1,3$ and $j=2,4$. In contrast to
our definition (\ref{P}) above, which has been
based on physically measurable quantities and which
involves a time restriction $\tau_1 \ll \tau_I$, the
definition (\ref{P_But}) is a time independent
correlator but probably cannot be measured in a
mesoscopic setting.
The idea put forward in Ref.\ \onlinecite{ssb_04}
then is, to use quantum mechanics to reexpress this
theoretical definition through measurable quantities.
In the tunneling limit, after transformation to electron-hole
currents, the current product term in (\ref{P_But}) can
be dropped (after proper experimental check, see the
discussion above) and the irreducible current correlator
can be expressed through the zero-frequency noise
correlator via
\begin{equation}
   S_{ij}^\mathrm{he} = \int dt
   \langle \delta \hat{I}^\mathrm{h}_i(t)
   \delta \hat{I}^\mathrm{e}_j(0) \rangle
   = \langle \delta \hat{I}^\mathrm{h}_i(0)
   \delta \hat{I}^\mathrm{e}_j(0) \rangle
   \tau_c
   \label{S_II}
\end{equation}
with the correlation time $\tau_c$ defined via
\begin{equation}
   \tau_c \equiv \int dt
   \frac{\langle \delta \hat{I}^\mathrm{h}_i(t)
   \delta \hat{I}^\mathrm{e}_j(0)\rangle}
   {\langle \delta \hat{I}^\mathrm{h}_i(0)
   \delta \hat{I}^\mathrm{e}_j(0)\rangle}.
   \label{tauc}
\end{equation}
Assuming that $\tau_c$ does not depend on the
lead indices $i$ and $j$, this time scale disappears
after normalization and one arrives at the formula
(\ref{bES}) expressed through the zero frequency
noise $S_{ij}^\mathrm{he}$ as derived above starting
from the measurable expression (\ref{P})
for the coincidence probability.

However, away from the tunneling limit, the current
product term in (\ref{P_But}) cannot be dropped and the
correlation time $\tau_c$ does not vanish any longer;
the proposal made in Ref.\ \onlinecite{ssb_04} then is to
construct the coincidence probability from the
combination
\begin{equation}
   P_{ij} \propto \tau_c^{-1} S_{ij}^\mathrm{ee}
   + \langle \hat{I}^\mathrm{e}_i\rangle
   \langle \hat{I}^\mathrm{e}_j \rangle
   \label{P_SII}
\end{equation}
which requires the measurement of the average currents
$\langle \hat{I}^\mathrm{e}_i\rangle$ and $\langle
\hat{I}^\mathrm{e}_j\rangle$ in addition to the
zero-frequency noise correlator. The problematic step
in this construction is the need for the {\it precise
quantitative} knowledge of the correlation time
$\tau_c$, since this parameter now is part of the
evaluation of the coincidence probability itself
(this is different from the above discussion of
the tunneling limit where a rough knowledge of
$\tau_c$ was sufficient in order to verify that
the current product term can be dropped from
(\ref{CNNzfn})). The appearance of this additional
parameter is a consequence of expressing the
equal time correlator
$\langle \delta\hat{I}^\mathrm{e}_i(t) \delta
\hat{I}^\mathrm{e}_j(t)\rangle = \int (d\omega/2\pi)
S_{ij}^\mathrm{ee}(\omega)$, involving all frequencies,
through the zero-frequency value $S_{ij}^\mathrm{ee}
(\omega=0)$ alone. The point of view put forward
in Ref.\ \onlinecite{ssb_04} then is that the
parameter $\tau_c$ shall be obtained from a
theoretical calculation, e.g., $\tau_c = 4\pi\tau_V$
in a non-interacting system where the irreducible
current correlator decays $\propto \sin^2(eVt/2\hbar)
/(eVt/2\hbar)^2$ (note that in Ref.\ \onlinecite{ssb_04}
this result is used in the expression for the
coincidence probability $P_{ij}$).
The proposal to replace $\tau_c$ in (\ref{P_SII})
through a theoretically calculated quantity avoids
the need for a short time measurement; on the other
hand, one has to accept that the coincidence
probability  obtained in this manner is subject to
the approximations (such as neglecting effects of
interactions, resonances, etc. present in the real
experiment) made in the theoretical evaluation
of $\tau_c$. Alternatively, one might want to
obtain $\tau_c$, cf.\ (\ref{tauc}), directly from
an experiment; however, as $\tau_c$ now is used
in the evaluation of the coincidence probability,
a precise knowledge of this parameter is required
and hence an accurate measurement of the current
correlator has to be performed. This then boils
down to a short time measurement and nothing can
be gained.

Below, we take the point of view that the Bell inequalities
should be built from physically measurable quantities.
Starting from the expression (\ref{E}), we proceed
with its theoretical evaluation in order to predict
the expected outcome of such a Bell inequality test within
a quantum mechanical frame. We first determine the finite
time current cross-correlator between leads `u' and `d'
for a stream of spinless fermions; the generalization to
the spin-1/2 case is straightforward. We express the BI
in terms of these finite time correlators and find its
violation for the general case expressed through electronic
correlators and for the tunneling case involving
electron-hole correlators. We also derive the results
expected from the alternative formulation based on
coincidence probabilities.

\section{Current cross-correlators}

\subsection{Bell inequalities with electron currents}

Starting from the field operators $\hat\Psi_\mathrm{u}$ and
$\hat\Psi_\mathrm{d}$ describing the scattering states in the
leads, we determine the (electronic) irreducible current
cross correlator $C^\mathrm{ee}_{x,y}(\tau) \equiv
\langle \delta \hat{I}^\mathrm{e}(\tau,x)
\delta\hat{I}^\mathrm{e}(0,y)\rangle$ with positions
$x$ and $y$ in the leads `u' and `d' using the
standard scattering theory of noise \cite{noise} and
split the result into an equilibrium component
$C^\mathrm{eq}_{x,y}(\tau;V=0)$ and an excess part
$C^\mathrm{ex}_{x,y}(\tau;V)$
\begin{eqnarray}
   &&C^\mathrm{eq}_{x,y}(\tau)
   =\frac{e^2T_\mathrm{du}}{h^2}
   \left[\alpha(\tau+\tau_+,\theta) +
   \alpha(\tau-\tau_+,\theta)\right],
   \label{Ceqex}\\
   &&C^\mathrm{ex}_{x,y}(\tau)
   =-\frac{4e^2T_\mathrm{u} T_\mathrm{d}}{h^2}
   \sin^2\frac{eV(\tau-\tau_-)}{2\hbar}
   \,\alpha(\tau-\tau_-,\theta),
   \nonumber
\end{eqnarray}
with $\alpha(\tau, \theta) = \pi^2\theta^2/\sinh^2 [\pi\theta\tau
/\hbar]$, $\tau^{\pm}= (x\pm y)/v_{\rm\scriptscriptstyle F}$, and
$\theta$ the temperature of the electronic reservoirs. In order to
arrive at the result (\ref{Ceqex}) we have dropped terms
\cite{noise} small in the parameter $|\epsilon^\prime-\epsilon|
/\epsilon_{\rm \scriptscriptstyle F}$ and have used the standard
reparametrization of the scattering matrix for a three-terminal
splitter (see Lesovik {\it et al.} in Ref.\ \onlinecite{ent_sc}).

Extending the above results to spin-1/2 particles, we introduce
the spin-projected field operators $\hat\Psi_{1,3} = \sum_\sigma
\langle \pm {\bf a}|\sigma\rangle \hat{\Psi}_{\mathrm{u}\sigma}$
(and similar for the `d' lead). The correlators
$C^\mathrm{ee}_{ij} ({\bf a},{\bf b})$ relate to the result
(\ref{Ceqex}) for spinless particles via $C^\mathrm{ee}_{ij}({\bf
a},{\bf b};\tau)= |\langle{\bf a}_i|{\bf b}_j\rangle|^2
C^\mathrm{ee}_{x,y} (\tau)$ with ${\bf a}_{1,3} = \pm{\bf a}$ and
${\bf b}_{2,4} = \pm{\bf b}$. The spin-projections derive from the
angle $\theta_{{\bf a}{\bf b}}$ between the directions ${\bf a}$
and ${\bf b}$ via $\langle \pm {\bf a}|\pm {\bf b} \rangle =
\cos^2 (\theta_{{\bf a}{\bf b}}/2)$ and $\langle \pm {\bf a}|\mp
{\bf b} \rangle = \sin^2 (\theta_{{\bf a}{\bf b}}/2)$ and the BI
(\ref{BI}) assumes the form
\begin{equation}
   \Bigg|\frac{C^\mathrm{ee}_{x,y}(\tau)
   [\cos\theta_{{\bf a}{\bf b}}
   \!-\!\cos\theta_{{\bf a}\bar{\bf b}}\!
   +\!\cos\theta_{\bar{\bf a}{\bf b}}
   \!+\!\cos\theta_{\bar{\bf a}\bar{\bf b}}]}
   {2 C^\mathrm{ee}_{x,y}(\tau)+
   \langle \hat{I}^\mathrm{e}_\mathrm{u}\rangle \langle
   \hat{I}^\mathrm{e}_\mathrm{d}\rangle}\Bigg| \!\leq  1.
   \label{BI_gen}
\end{equation}
Its maximum violation is obtained for the standard orientations
of the detector polarizations, $\theta_{{\bf a}{\bf
b}}=\theta_{\bar{\bf a}{\bf b}} =\theta_{\bar{\bf a}\bar{\bf
b}}=\pi/4$, $\theta_{{\bf a} \bar{\bf b}}=3\pi/4$ and the BI
reduces to
\begin{equation}
   E^\mathrm{ee}_{\rm\scriptscriptstyle BI} \equiv
   \left|\frac{2C^\mathrm{ee}_{x,y}(\tau)}{
   2C^\mathrm{ee}_{x,y}(\tau)
   +\langle \hat{I}^\mathrm{e}_\mathrm{u}\rangle\langle
   \hat{I}^\mathrm{e}_\mathrm{d}\rangle} \right|
   \leq\frac1{\sqrt{2}}.
   \label{BI_max}
\end{equation}
In the limit of low temperatures $\theta < eV$ and for large
distances $x = y \gg \tau_V v_{\rm\scriptscriptstyle F}$ (allowing
us to neglect the equilibrium part in the correlator
$C^\mathrm{ee}_{x,y}$) the above expression (\ref{BI_max}) reduces
to the particularly simple form
\begin{equation}
   \frac{2\sin^2(eV\tau/2\hbar)}{\tau^2(eV/\hbar)^2
   -2\sin^2(eV\tau/2\hbar)} \leq \frac1{\sqrt{2}},
\label{Bell_simple}
\end{equation}
where we have used that $\langle \hat{I}^\mathrm{e}_\mathrm{u}
\rangle = (2e^2/h) T_\mathrm{u} V$ and $\langle
\hat{I}^\mathrm{e}_\mathrm{d}\rangle = (2e^2/h) T_\mathrm{d} V$. We
observe that in this limit the Bell inequality is {\it i)}
violated at short times $\tau < \tau_{\rm\scriptscriptstyle BI} =
\tau_V$, see Fig.\ 2, {\it ii)} this violation is independent of
the transparencies $T_\mathrm{u}$, $T_\mathrm{d}$ and hence
universal, and {\it iii)} the product of average currents
$\Lambda^\mathrm{ee}_+$ is the largest term in the denominator
of (\ref{BI_gen}) and hence always relevant. Note that the
important quantity appearing in (\ref{BI_max}) is the
space and time-dependent correlator $C^\mathrm{ee}_{x,y}(\tau)$.
The small quantity required for the violation of the
BI then is the shifted time $\tau-\tau^- < \tau_V$;
placing the detectors a finite distance apart one may
make use of the additional time {\it delay}, although
the time {\it resolution} remains unchanged.

In the low frequency analysis of Ref.\
\onlinecite{chtchelkatchev_02} no violation in the BI
had been found for a normal injector, in agreement with
the results found here. On the other hand, it has been
realized in Ref.\ \onlinecite{beenakker_03b} that a
short time measurement on the scale $\tau_V$ can
exhibit entanglement in a normal system. In particular,
the proper use of the Bell parameter $\tilde{E}_{\rm
\scriptscriptstyle BI}$ with $\tilde{E}$ given by
(\ref{bE}) provided such an entanglement away from
the tunneling limit, while the use of the expression
(\ref{bES}) led to a `spurious amplification factor' in
the relation between the Bell parameter and the concurrence.
\begin{figure} [h]
   \includegraphics[scale=0.450]{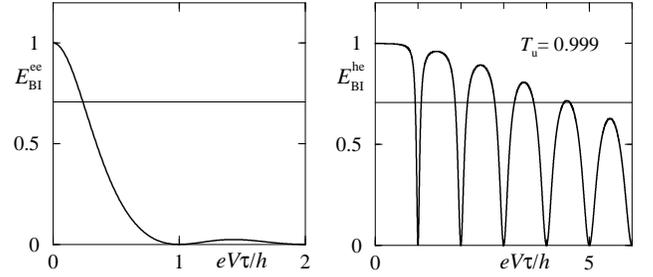}
   \caption[]{Bell inequality test for electron-electron
   (left) and electron-hole (right) currents. The thin
   line at $1/\sqrt{2}$ marks the critical value above
   which the Bell inequality is violated.}
   \label{fig:BIeh}
\end{figure}

\subsection{Bell inequalities with electron-hole currents;
tunneling limit}

Next, we consider the tunneling limit and determine the outcome of
a Bell measurement involving a hole current
$\hat{I}^\mathrm{h}_\mathrm{u} \equiv (2e^2/h) V -
\hat{I}^\mathrm{e}_\mathrm{u}$ in the `up' lead and the electronic
current $\hat{I}^\mathrm{e}_\mathrm{d}$ in the `down' lead.
The cross-measurement in different leads implies that the setup
is sensitive only to the split-pair part of the scattering wave
function $|\Psi_\mathrm{out}^\mathrm{he} \rangle$ which is fully
spin-entangled and hence the Bell inequality can be maximally
violated. In the tunneling limit, the correction term $\propto
t_\mathrm{sd}^2$ also contributes a signal and spoils the maximal
violation, ultimately limiting the use of the tunneling limit
to devices with large enough transparency $T_\mathrm{u}$ in
the well conducting lead (small enough transparency
$T_\mathrm{d}$ in the blocked lead).

The calculation proceeds as above but now involves the electron-hole
correlator $C^\mathrm{he}_{x,y}(\tau) \equiv \langle \delta
\hat{I}^\mathrm{h}_\mathrm{u}(\tau,x) \delta
\hat{I}^\mathrm{e}_\mathrm{d}(0,y)\rangle =
-C^\mathrm{ee}_{x,y}(\tau)$ and the product of the electron and
hole currents $\Lambda^\mathrm{he}_+ = \langle
\hat{I}^\mathrm{h}_\mathrm{u} \rangle \langle
\hat{I}^\mathrm{e}_\mathrm{d}\rangle = (2e^2/h)^2 T_\mathrm{d}
(1-T_\mathrm{u}) \, V^2$ (again, $\Lambda^\mathrm{he}_- =0$).
The Bell inequality corresponding to (\ref{Bell_simple}) now reads
\[
      E^\mathrm{he}_{\rm\scriptscriptstyle BI}
      \equiv
      \frac{2\sin^2(eV\tau/2\hbar)}{
      2\sin^2(eV\tau/2\hbar)+T_\mathrm{u}^{-1}
      (1-T_\mathrm{u})\tau^2 (eV/\hbar)^2}
      \leq\frac1{\sqrt{2}}
\]
and an illustration of this result is given in Fig.\
\ref{fig:BIeh}. The violation of the Bell inequality in the
tunneling limit exhibits a much richer structure: {\it i)} the
violation requires a minimum transparency $T_\mathrm{u}$ in the
upper lead: evaluating $E^\mathrm{he}_{\rm\scriptscriptstyle BI}$
at $\tau=0^+$, we obtain the condition $T_\mathrm{u} >
T_\mathrm{min} \equiv 2/(\sqrt{2}+1) \approx 0.83$. {\it ii)} For
$T_\mathrm{u}> T_\mathrm{min}$ the Bell inequality is violated
during times $\tau< \tau_{\rm\scriptscriptstyle BI} = \tau_V
\sqrt{2-\sqrt{2}}\, \sqrt{T_\mathrm{u} (1-T_\mathrm{u})^{-1}}
\approx \tau_V/ \sqrt{T_\mathrm{d}}$, where we have assumed
$1-T_\mathrm{u} \approx T_\mathrm{d}$ as is the case for a
splitter with a small back reflection $r_\mathrm{s} \ll 1$;
this result is different from the time limitation noted
in Ref.\ \onlinecite{beenakker_03b}. {\it
iii)} The BI remains un-vio\-lated in narrow intermediate regions
separated by the single particle correlation time $\tau_V$ and
decreases slowly $\propto \tau^{-2}$ with increasing time. {\it
iv)} The product of average currents $\Lambda^\mathrm{he}_+$ gives
a small correction to the denominator in $E^\mathrm{he}_{\rm
\scriptscriptstyle BI}$ at short times. The comparison with the
electronic result is quite striking: the time interval over which
the Bell inequality is violated is extended by a factor
$1/\sqrt{T_\mathrm{d}} \gg 1$ and the universality (i.e., the
independence on the transmissions $T_\mathrm{u}$ and
$T_\mathrm{d}$) is lost.

\subsection{Bell inequalities with number correlators}

Finally, we quote the results obtained for the Bell
measurement based on coincidence probabilities or
number correlators, cf.\ Eq.\ (\ref{bE}) and Refs.\
\onlinecite{chtchelkatchev_02,beenakker_03b}.
Again, we split the number correlator $K^\mathrm{ee}_{ij}$
(cf.\ (\ref{Kdef}), we use the electronic version)
into an orbital- and a spin component,
$K^\mathrm{ee}_{ij} = |\langle {\bf a}_i|{\bf b}_j
\rangle|^2 K_{x,y}^\mathrm{ee} (\tau)$ and assume the
standard set of directions ${\bf a}$, ${\bf b}$,
$\bar{\bf a}$, and $\bar{\bf b}$ in order to arrive
at the inequality
\begin{equation}
   \Bigg|\frac{2K^\mathrm{ee}_{x,y}(\tau)}
   {2K_{x,y}^\mathrm{ee} (\tau) + \tau^2
   \langle\hat I^\mathrm{e}_\mathrm{u}\rangle\langle
   \hat I^\mathrm{e}_\mathrm{d}\rangle}\Bigg|
   \leq\frac1{\sqrt{2}}.
   \label{BI_K_ee}
\end{equation}
Again, this electronic Bell inequality is universal
(cf.\ Eq.\ (\ref{Ceqex})) and violated at short times
$\tau < \tau_V$, cf.\ also Ref.\ \onlinecite{beenakker_03b}.

The tunneling limit involving the electron-hole
number correlator $K_{x,y}^\mathrm{he}(\tau)=
\langle \delta \hat{N}_\mathrm{u}^\mathrm{h}(x,\tau)
\delta \hat{N}_\mathrm{d}^\mathrm{e}(y,\tau)\rangle$
is more interesting: The Bell inequality
takes the form (\ref{BI_K_ee}) but with $K^\mathrm{ee}$ and
$\langle\hat I^\mathrm{e}_\mathrm{u}\rangle\langle \hat
I^\mathrm{e}_\mathrm{d}\rangle$ replaced by $K^\mathrm{he}$ and
$\langle\hat I^\mathrm{h}_\mathrm{u}\rangle\langle \hat
I^\mathrm{e}_\mathrm{d}\rangle$. Its evaluation at $\tau = 0^+$
provides the same condition $T_\mathrm{u} > T_\mathrm{min}$ as
found previously for the violation of the Bell inequality. In the
tunneling limit $(1-T_\mathrm{u})/T_\mathrm{u}\ll 1$, the number
correlator can be estimated at large times $\tau \gg \tau_V$ and
we find  $K_{x,y}^\mathrm{he}(\tau)\approx (e^2/2\pi) T_\mathrm{u}
T_\mathrm{d}(eV\tau/\hbar)$; the Bell inequality then is
violated for even larger times $\tau<\tau_{\rm\scriptscriptstyle
BI}=\tau_V(\sqrt{2}-1)\pi T_\mathrm{u}(1-T_\mathrm{u})^{-1}
\approx\tau_V/T_\mathrm{d}$. Hence we see that the violation
appears just over those time scales $\tau < \tau_I$ where the
number correlator ${\cal K}^\mathrm{he}_{ij}(\tau)$ assumes the
meaning of a coincidence probability. Note that the time
dependence found here is lost once we drop the current product
term $\tau^2 \langle\hat I^\mathrm{h}_\mathrm{u}\rangle\langle
\hat I^\mathrm{e}_\mathrm{d}\rangle$, taking us to the time
independent result corresponding to (\ref{bES}).

\subsection{Origin of entanglement}

An interesting question concerns the origin of the
entanglement detected in the Bell inequality measurement
described in the present paper. We note that in
Refs.\ \onlinecite{beenakker_03,beenakker_03b}, the
entanglement had been attributed to the elastic scattering
in the Fermi sea, although the proper selection of a projected
wave function component during the calculation corresponds
to a post-selection. In Ref.\ \onlinecite{ssb_04} post-selection
was noted to be the origin of entanglement; such post-selection
creating entanglement is a well known mechanism in optics
\cite{shih_88}. In both of the above cases, the entangled
degrees of freedom originated from independent reservoirs.
The situation is slightly more complicated in the present
case: As shown above, our setup involves a simple normal
reservoir injecting spin-singlet correlated pairs of electrons
into the source lead which are conveniently described by the
wave function $\Psi_\mathrm{in}^{\scriptscriptstyle 12}
= \phi_\mathrm{s}^{\scriptscriptstyle 1}
\phi_\mathrm{s}^{\scriptscriptstyle 2}
\chi_\mathrm{sg}^{\scriptscriptstyle 12}$. These {\it local}
spin-singlet pairs are subsequently separated in space
by a beam splitter and detected in a coincidence
measurement. The measurement is only sensitive to pairs
of particles propagating in different arms, implying a
post-selection or projection of the scattering wave function
during which only its cross term describing a {\it split}
spin-singlet pair survives. In this context it is interesting
to note that the incoming local spin-singlet,
being a simple Slater determinant, is not entangled
according to the definition given by Schliemann {\it et al.}
\cite{schliemann_01}  However, after the beam splitter
the orbital wave function $\phi_\mathrm{s}$ is delocalized
between the two leads, $\phi_\mathrm{s} \rightarrow \Phi
= t_\mathrm{su}\phi_\mathrm{u}+t_\mathrm{sd}\phi_\mathrm{d}$.
While the scattered state remains a Slater
determinant $\Psi_\mathrm{out}^{\scriptscriptstyle 12}
= \Phi^{\scriptscriptstyle 1} \Phi^{\scriptscriptstyle 2}
\chi_\mathrm{sg}^{\scriptscriptstyle 12}$, the singlet
correlations now can be observed in a coincidence
measurement testing the cross-correlations between the
leads `u' and `d'. Hence the original spin-entanglement is
produced by the reservoir, but its observation requires proper
projection. It is then difficult to trace a unique origin
for the entanglement manifested in the present Bell
inequality test. An appropriate setup addressing this
question should involve independent reservoirs injecting
the particles carrying the degrees of freedom to be
entangled, e.g., particles with opposite spin residing
in a Slater determinant of the form
$\Psi_\mathrm{in}^{\scriptscriptstyle 12}
= [\phi_{\mathrm{s}\uparrow}^{\scriptscriptstyle 1}
\bar{\phi}_{\mathrm{s}\downarrow}^{\scriptscriptstyle 2}
- \bar{\phi}_{\mathrm{s}\downarrow}^{\scriptscriptstyle 1}
\phi_{\mathrm{s}\uparrow}^{\scriptscriptstyle 2}]/\sqrt{2}$
which is not entangled in the spin variable. Such an analysis
has been presented in Ref.\ \onlinecite{andrei_04}
with the result, that the orbital projection in the coincidence
measurement is sufficient to produce a spin-entangled state.

\section{Conclusion}

In conclusion, we find that spin-entangled pairs of electrons
exist and leave their trace in the violation of Bell inequalities
in a mesoscopic setup even in the absence of interaction. The
source of entanglement is traced back to the nature of injected
electrons forming a regular stream of locally singlet-correlated
particles combined with a post-selection \cite{shih_88,ssb_04}
during the Bell type measurement. The splitter itself does
not contribute to the entanglement of the pair, but fulfills
the crucial task of separating the spin-entangled constituents
of the pair in real space, thus rendering them useful as a
quantum resource of entanglement. While most of the previous
analysis of entanglement was restricted to the tunneling limit
\cite{chtchelkatchev_02,samuelsson_03,beenakker_03}, we have
overcome this restriction and have demonstrated universal
violation of BIs in setups based on electron correlators.
We have determined the degree and duration in time of the
BI violation and have found pronounced dependencies on the
choice of observables.

We thank Carlo Beenakker and Markus B\"uttiker for discussions
and acknowledge financial
support from the Swiss National Foundation (SCOPES and CTS-ETHZ),
the Forschungszentrum J\"ulich within the framework of the Landau
Program, the Russian Science Support Foundation, the Russian
Dynasty Foundation, the Russian Ministry of Science, and the
program `Quantum Macrophysics' of the RAS.


\begin{thebibliography}{99}
\bibitem{note} Closer inspection of Ref.\ \onlinecite{samuelsson_03}
actually reveals that the orbital entanglement is not
due to the attractive interaction in the superconducting
injectors; the latter only serve as injectors of pairs
and the entanglement is also produced without the presence
of interaction.
\end{thebibliography}
\end{document}